\documentclass[aps,twocolumn,secnumarabic,nobalancelastpage,amsmath,amssymb,
nofootinbib]{revtex4-1}

\usepackage{graphicx}      % alternative graphics specifications
\usepackage{longtable}     % helps with long table options
\usepackage{url}           % for on-line citations
\usepackage{bm}            % special 'bold-math' package
\usepackage{amsmath}
\usepackage{epstopdf}
\begin{document}
\title{Opinion diversity and community formation in adaptive networks}

\author{Yi Yu$^{1}$}
\author{Gaoxi Xiao$^{2,3}$}
  \email{Email address: egxxiao@ntu.edu.sg}
\author{Guoqi Li$^{4}$}
\author{Wee Peng Tay$^{2}$}
\author{Hao Fatt Teoh$^{1}$}
\affiliation{
$^{1}$ Data Science Innovation Hub, Merck Sharp \& Dohme, 1 Fusionopolis Way, Singapore 138632\\
$^{2}$ School of Electrical and Electronic Engineering, Nanyang Technological University, 50 Nanyang Avenue, Singapore 639798\\
$^{3}$ Complexity Institute, Nanyang Technological University, 50 Nanyang Avenue, Singapore\\
$^{4}$ Center for Brain Inspired Computing Research, Department of Precision Instrument, Tsinghua University, Beijing, China\\
}

\date{\today}

\begin{abstract}
It is interesting and of significant importance to investigate how network structures co-evolve with opinions. The existing models of such co-evolution typically predict that network nodes either reach a global consensus or break into separated communities, each of which has its own community consensus. Such results, however, cannot explain the richness of real-life opinions, which are typically diversified with no global or even community consensus. In addition, people seldom, if not never, totally cut themselves off from dissenters. In this article, we show that, a simple model integrating consensus formation, link rewiring and opinion change allows complex system dynamics to emerge, driving the system into a dynamic equilibrium with co-existence of diversified opinions. Specifically, similar opinion holders may form into communities yet with no strict community consensus; and rather than being separated into disconnected communities, different communities are connected by a non-trivial proportion of inter-community links. More importantly, we show that the complex dynamics may lead to different numbers of communities at steady state with a given tolerance between different opinion holders. We construct a framework for theoretically analyzing the co-evolution process. Theoretical analysis and extensive simulation results reveal some useful insights into the complex co-evolution process, including the formation of dynamic equilibrium, the phase transition between different steady states with different numbers of communities, and the dynamics between opinion distribution and network modularity.
\end{abstract}

\maketitle

%%%%%%%%%%%%%%%%%%%%%%%%%%%%%%%%%%%%%%%%%%%%%%%%%%%%%%%%%%%%%%%%%%%%%%%%%%%%%
\section{Introduction}\label{sec:introduction}
Real-life adaptive complex networks co-evolve with the opinions, strategies and actions of the individuals in the network.  For example, upon the outbreak of a dangerous infectious disease, people tend to avoid contact with those who are infected, which changes the contact network topology and consequently changes the dynamics of epidemic spreading \cite{1,2,3,4,5,6}. In social networks, generally speaking, people tend to make friends with those who share similar opinions and such friendship may in turn help strengthen their common beliefs \cite{7,8}. Other examples include the co-evolution of languages and social structures \cite{9,10,11,12}, and the co-evolution of modern cities and urban transportation systems \cite{13,14}.

Extensive studies have been carried out to investigate the co-evolution of network structure and opinions \cite{7,8,15,16,17,18,19,20,21}; the earliest of which may be the one by Holme and Newman \cite{7}. Based on the well-known voter model where a randomly chosen node could persuade a random neighbor to adopt the same opinion \cite{15,16}, they further took into account the evolution of network topology by introducing a simple rewiring mechanism, where nodes have a certain chance to rewire their links with dissenters to similar opinion holders (when the difference between two nodes' opinions is larger than a certain tolerance level, these two nodes are said to be dissenters to each other). It is found that there exists a phase transition in the change of rewiring rate: the final state of the system varies from reaching global consensus to breaking into separated communities, each of which reaching its own community consensus. Follow-up studies include deriving analytical solutions for the phase transition \cite{17,18}, extending the model to directed networks \cite{19}, and introducing opinion noises \cite{20} and different interaction mechanisms such as connecting nodes with different opinions \cite{21} and self-interaction \cite{22}, etc.

Another popular model for studying the interactions and dynamics between different opinions is the Deffuant model \cite{23,24,25,26,27}, where there is a continuous distribution of different opinions and a randomly chosen node may make consensus with a randomly selected neighbor holding similar opinion; their opinions hence come closer to each other or become the same. Similar to what happened to the voter model, rewiring was later introduced into the Deffuant model \cite{8}. It is shown that the existence of rewiring makes it harder for an adaptive network to reach global consensus; network may finally evolve into a few big opinion communities, each of which holding its community consensus.

Opinion change was introduced in \cite{28,29,30}, termed as ``noises" in these references. It is shown that in the Deffuant model with a continuous opinion distribution and a fixed network topology, different speeds of random opinion change may drive the system either to an ordered state with a set of well-defined opinion groups, or a disordered state where the opinion distribution tends to be uniform.

It can be observed that most of these existing models lead to similar final steady states where the network either reaches a global consensus or breaks into disconnected communities, each of which reaches its own consensus. Such observations however do not reflect common reality: i) real-world opinions are typically widely diversified with no global consensus \cite{40, 41}; and ii) people seldom get totally cut off from dissenters even when they want to do so (Arguably, it may be claimed that in human history a few extremism regimes tried to do so yet have all failed.). ``Mainstream opinions" may exist, but they do not become global consensus eliminating other opinions; and mainstream opinions themselves evolve, sometimes rather quickly.  While it is understood that many details have to be omitted to allow mathematical modeling focus on the most important factors, the fact that the existing models lead to steady state essentially different from real-life observations nevertheless calls for closer examination of these models. The questions include what factors should be included in the modeling to allow the emergence of a system steady state with ``real-life features" as stated above, and how these factors contribute to defining the steady state of the systems, etc.

In this article, we make a critical extension to the adaptive co-evolution model. For the first time to the best of our knowledge, we allow three important factors to be integrated into the same model, namely i) consensus formation, where directly connected network nodes may try to reach consensus if their opinions are similar to each other; ii) link rewiring, where network nodes may rewire some of their links with dissenters to similar opinion holders; and iii) mutation, where network nodes may change their opinions for various reasons other than consensus making. Examples of opinion mutation include change of mind due to an unexpected critical event, religion conversion \cite{31,32,33}, change in opinion due to abruptly changed environment (e.g., immigration \cite{34}), etc. In such cases, an individual's opinion may change quickly and significantly, while still keeping a significant part of his/her original connections \cite{35}. These individuals thus may share a similar opinion with one community while being connected to another community. They hence may become the "bridges" between different communities with significantly different opinions. We adopt the continuous opinion model where an opinion is generated by a continuous distribution over the region [0, 1]. Note that we define an opinion community as a group of opinion holders holding either the same opinion (when there is no mutation) or similar opinions with a bell-curve shape distribution  (when mutation exists) and with relatively denser connections in between. As we will see in this article, with all the three factors being integrated into the same model, complex system dynamics emerge. Specifically, instead of being broken into separate communities each holding its own community consensus, similar opinions would form into communities interconnected by a non-trivial number of inter-community links. In each community, different opinions coexist, typically following a stable, bell-curve-style distribution. As discussed earlier, such a model may arguably better resemble what we may observe in real life, and it allows complex system dynamics to quickly emerge when intruders (e.g., new ideas) come in or when certain internal/external driving forces are applied to the underlying social networks. Interestingly, it is found that the integration of the three factors may also lead to different numbers of communities in the steady-state network with a given tolerance level: the existence of opinion mutation tends to increase the number of opinion communities.

We develop a theoretical framework to describe the co-evolution process. Theoretical analysis and extensive simulation results help reveal some useful insights into the system evolution process, including the formation of dynamic equilibrium, the transition between steady states with different number of communities, and the dynamics between opinion distribution and network modularity.

\section{adaptive model}\label{sec:model}
In this article, we adopt the continuous opinion model where an opinion is generated by a continuous distribution over the region $[0,1]$.  As discussed earlier, there are three key components integrated into this model, namely consensus formation, link rewiring and opinion mutation respectively. Specifically, each node in the network is initialized with a uniformly distributed random opinion value between 0 and 1. At each time step $t$, a node $A$ is randomly selected together with its random neighbor $B$. Denote the opinion of a node $X$ at time $t$ as $o(t,X)$ . If $|o(t,A)-o(t,B)|>d$, where $d$ is the tolerance threshold, with a probability $w$, node $A$ would rewire the link to a randomly selected node $C$ sharing similar opinion with node $A$, i.e., $|o(t,A)-o(t,C)|\le d$; if $|o(t,A)-o(t,B)|\le d$, then with a probability $1-w$, the two nodes would perform consensus by updating their opinions as follows:
\begin{equation}
\begin{split}
o(t+1,A)=o(t,A)-\mu[o(t,A)-o(t,B)];\\
o(t+1,B)=o(t,B)+\mu[o(t,A)-o(t,B)];
\end{split}
\end{equation}	
\begin{figure*}[!hbt]
\begin{center}
\includegraphics[width=160mm]{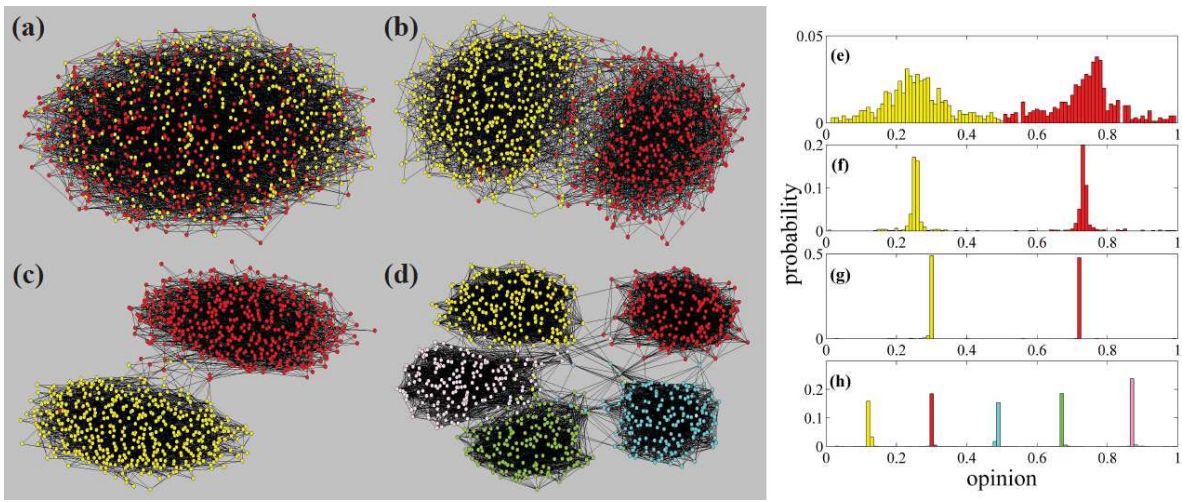}
\caption []{
\label{fig:fig1}{Network structures at the steady state where (a) $d=0.25$, $p=0.1$, $w=0.5$; (b) $d=0.25$, $p=0.01$, $w=0.5$; (c) $d=0.25$, $p=0.001$, $w=0.5$; and (d) $d=0.1$, $p=0.001$, $w=0.5$; and their corresponding opinion distribution where (e) $d=0.25$, $p=0.1$, $w=0.5$; (f) $d=0.25$, $p=0.01$, $w=0.5$; (g) $d=0.25$, $p=0.001$, $w=0.5$; and (h) $d=0.1$, $p=0.001$, $w=0.5$. The network starts as an ER random network with a size of $N=10^3$ and an average nodal degree of $\langle k\rangle=10$. For Figs. (a), (b) and (c), nodes in yellow and red respectively hold opinions within range of $[0, 1/2]$ and $(1/2, 1]$. For Fig. (d), nodes in yellow, red, blue, green and pink respectively hold opinions within the range of $[0, 1/5]$, $(1/5, 2/5]$, $(2/5, 3/5]$, $(3/5, 4/5]$ and $(4/5, 1]$.}}
\end{center}
\end{figure*}
where $\mu\in(0,1/2]$. Following most of the existing literature \cite{23,24,25,26,27,28,29,30}, for simplicity we let $\mu=1/2$ in the rest of the paper.
Further more, with a probability $p$, a node may change its opinion to another random value in $[0,1]$. We term such kind of opinion change as mutation and $p$ as the $mutation~rate$.

\section{Results}\label{sec:results}

In this section, we show that with the proposed co-evolution model, a network evolves into a dynamic equilibrium with network nodes forming into several communities connected by a non-trivial proportion of inter-community links; and the nodes in each community hold a certain range of opinions with a bell-curve-style distribution. We also show the dynamic process of the co-evolution between opinions and network topology and explain the formation of the dynamic equilibrium.

To start with some relatively simple examples, Fig. 1 shows the snapshots of some networks and their opinion distributions at steady state. Different values of mutation rate $p$ and tolerance threshold $d$ are adopted. The network starts as an ER random network \cite{35} with a size of $N=10^3$, an average nodal degree of $\langle k\rangle=10$, and a uniform distribution of different opinions. The simulation lasts for $2\times10^6$ steps, where at each step a single node is randomly selected, and with probability $1-w$, this node may make consensus with one of its neighbors or with probability $w$, to rewire one of its links. After that, a node may randomly mutate its own opinion at a probability $p$. Our simulations confirm that the number of time steps is large enough for the system to reach the steady state.

\begin{figure*}[!hbt]
\begin{center}
\includegraphics[width=150mm]{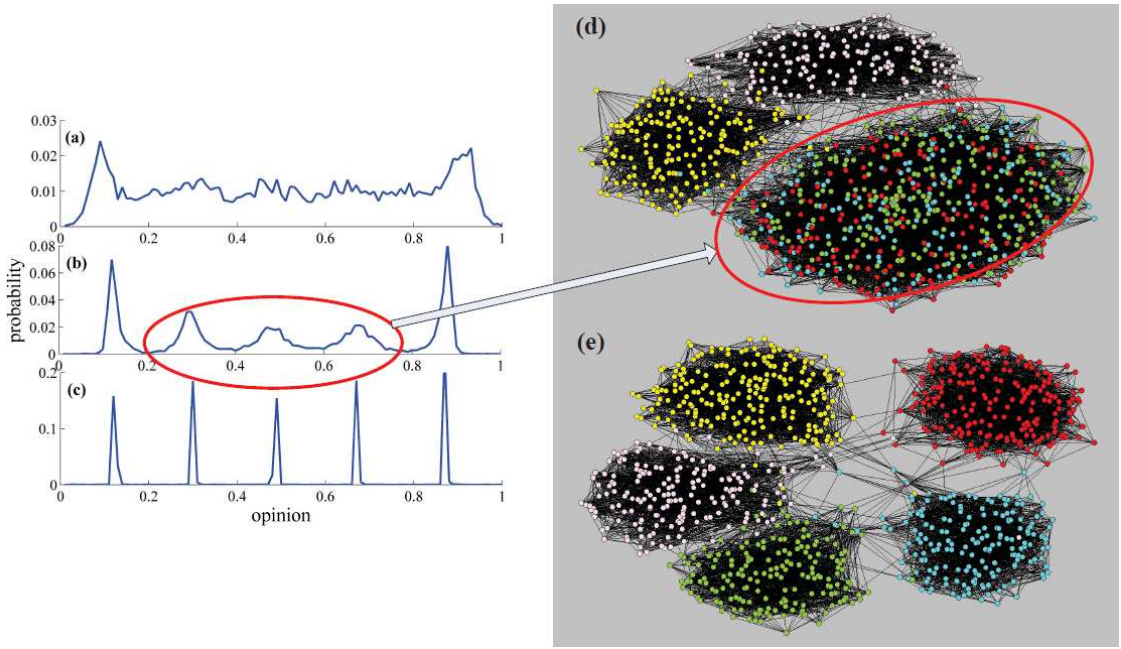}
\caption [An example of applying the MMI algorithm to allocate 2 monitors in an 8-node network.]{
\label{fig:fig2}{Opinion distribution at different time steps of the evolution process: (a) $t=500$, (b) $t=3000$, (c) $t=2\times10^6$; and the network topology at (d) $t=3000$ and (e) $t=2\times10^6$. $d=0.1$, $p=0.001$, and $w=0.5$. The network starts as an ER random network with a size of $N=10^3$ and an average nodal degree of $\langle k\rangle=10$. Every time step, a single pair of nodes is chosen.}}
\end{center}
\end{figure*}

Figs.~1(a) to 1(d) show the network topologies of 4 different cases with different values of $d$, $w$ and $p$, where nodes form into several communities connected by a non-trivial proportion of inter-community links. Figs.~1(e) to 1(h) show the corresponding distributions of nodes' opinions in these 4 cases, respectively. We see that in all the 4 cases, several bell-curve-style peaks with approximately equal ``heights" are formed, and a higher mutation rate leads to denser inter-community connections (comparing Figs.~1(a), 1(b) and 1(c)) and more diversified opinion distributions (comparing Figs.~1(e), 1(f) and 1(g)). Notice that the system reaches a dynamic equilibrium: nodes will keep making consensus, changing their opinions and constantly rewire their links. Nevertheless, the basic statistical properties such as opinion distribution and the proportion of inter-community links remain stable.

To further examine how nodes with different opinions distribute in different communities, we paint nodes with opinions in different ranges with different colors. Specifically, for a network finally forming into $M$ communities, nodes holding opinions within the interval $I_i$ (we define $I_1$ as $[0,1/M]$ and $I_i$ as $((i-1)/M,i/M]$ for $i=2,...,M$ respectively) are painted in the same color. For example, for networks in Figs.~1(a) to 1(c) where two communities are formed, nodes holding opinions within $[0, 0.5]$ and $(0.5, 1]$ are painted in yellow and red respectively. Examining Figs.~1(a) to 1(d), we find that nodes with the same color mostly stay in the same community in the final state. This means that nodes in the same community mostly share similar (but not necessarily the same) opinions. More specifically, the opinions of the nodes in the same community typically form into a bell curve with a peak value approximately at the center as the ``mainstream opinion". A theoretical framework and detailed analysis will be presented in Section 4.1.

Note that the existence of opinion mutation plays a critical role in defining the final state: existing studies \cite{7, 8, 15,16,17,18,19,20,21,22} have shown that the integration of consensus formation and link rewiring leads to steady state with a few disconnected communities, each holding its single-value community consensus. We now examine the process of the co-evolution between opinion and network topology which finally lead to the formation of the dynamic equilibrium. We examine the case where $d=0.1$, $p=0.001$, and $w=0.5$, which shall finally leads to the formation of 5 different communities. Figs.~2(a), 2(b) and 2(c) show the respective opinion distributions at early, middle and final stage of the co-evolution process; while Figs.~2(d) and 2(e) show the corresponding network topologies at the middle and final stages, respectively. We number the 5 communities from 1 to 5 in an increasing order of their respective middle opinion values and paint their nodes in yellow, red, light blue, green and pink colors respectively. The whole evolution process can be viewed as an instability propagating from the boundaries towards the center (Figs.~2(a), 2(b) and 2(c)). Specifically, at the early stage of the evolution, the two peaks of the opinion distribution close to the boundaries (i.e., opinions 0 and 1) form up (Fig.~2(a)) and quickly grow into bell-curve-style peaks (Fig.~2(b)). Looking at Fig.~2(d), we can see that at this moment, communities 1 and 5, which correspond to these two opinion peaks, have already grown mature and become largely separated from the rest of the network, while the rest part remains as a giant component mainly holding opinions ranging from 1/5 to 4/5 (indicated by the red circles in Figs.~2(b) and 2(d)). In the middle stage, communities 2 and 4 emerge from this giant component and then largely separate themselves from the other nodes. This process continues until the dynamic equilibrium is finally reached. Similar process has been observed for all the different values of $d$: formation of communities starts from the two communities holding opinions closest to the boundaries and then propagates inbound towards the central opinion value.

\begin{figure*}
\begin{minipage}{0.45\textwidth}
  \centerline{\includegraphics[width=8.0cm]{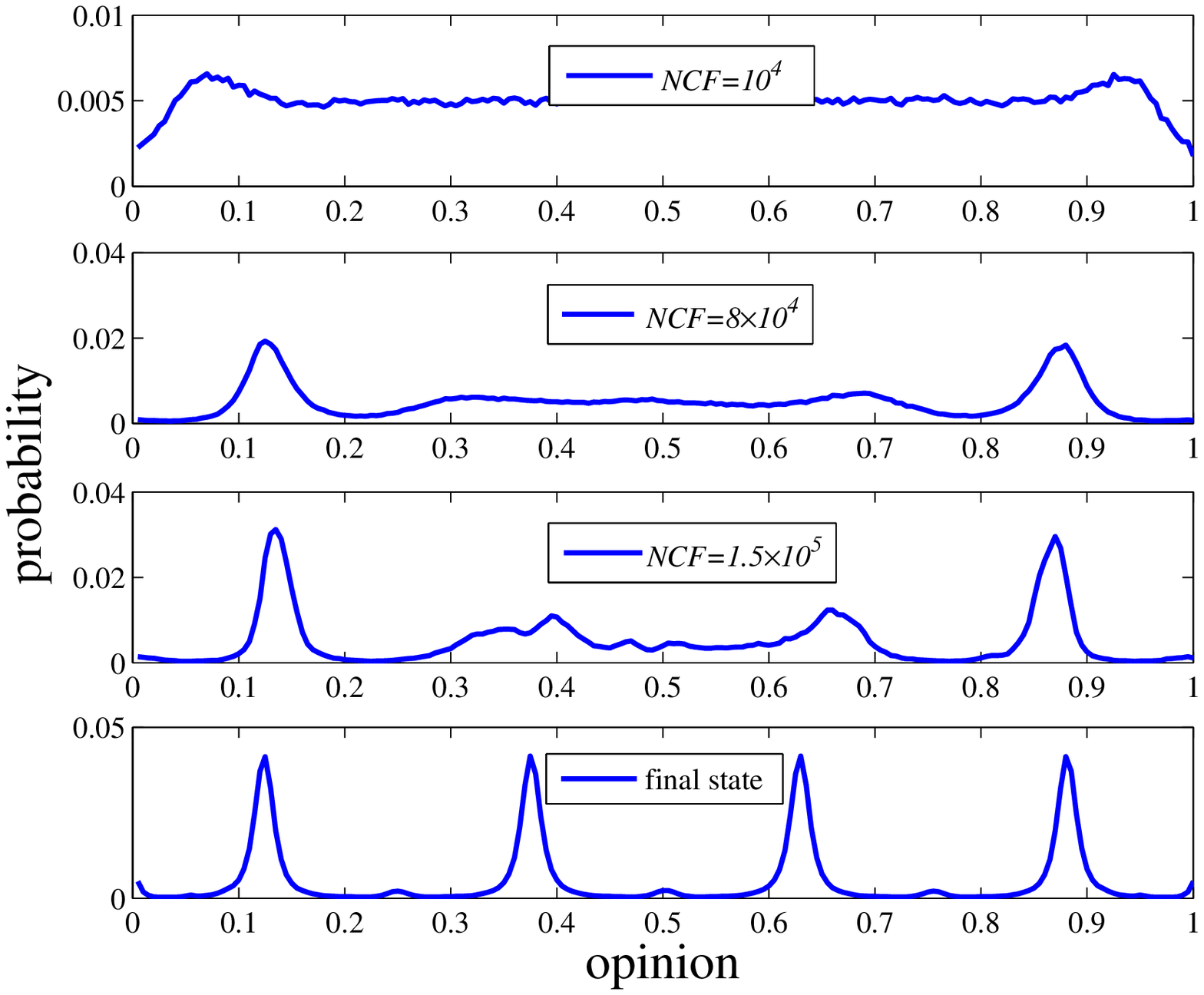}}
  \centerline{(a)}
\end{minipage}
\begin{minipage}{0.45\textwidth}
  \centerline{\includegraphics[width=8.0cm]{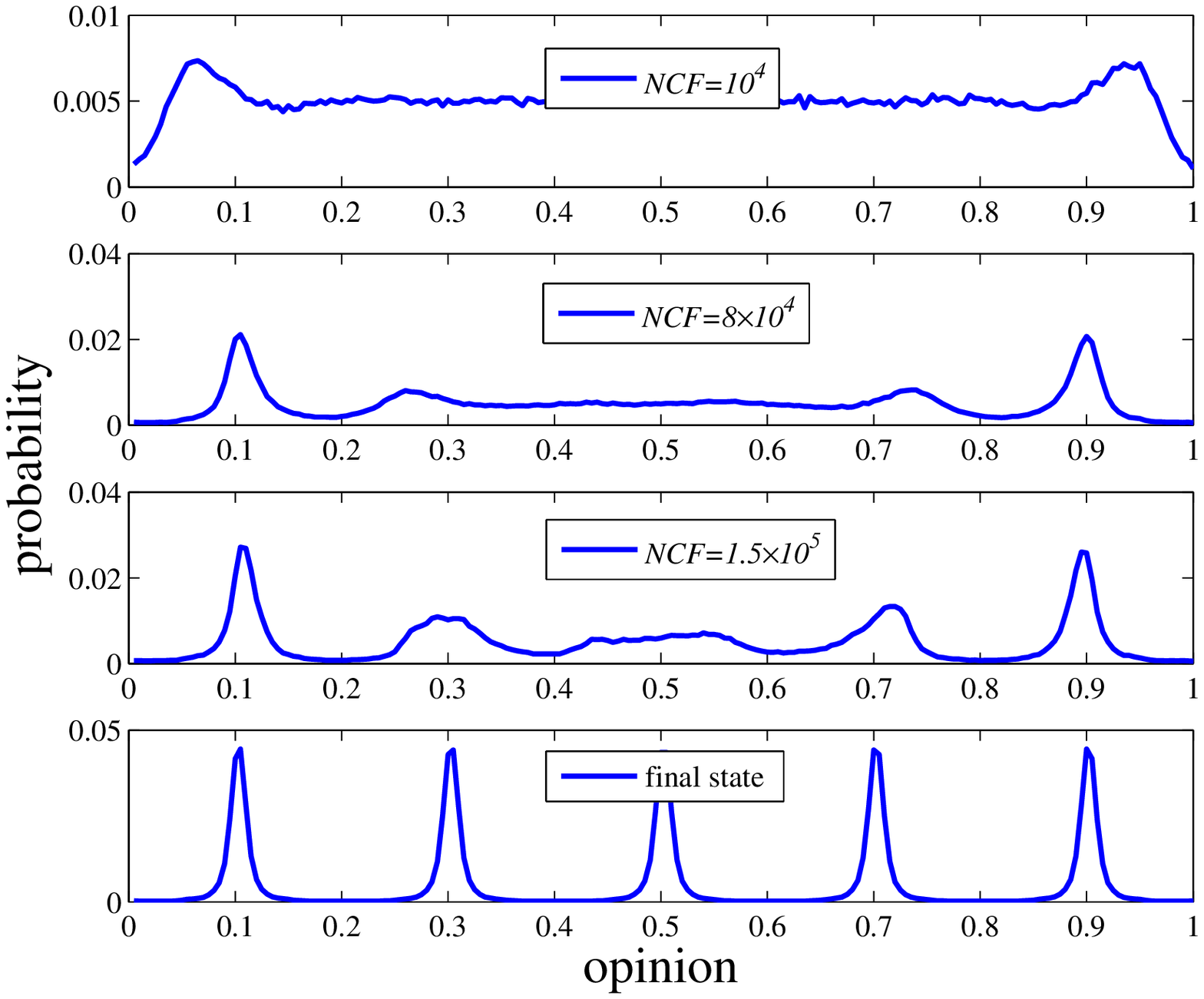}}
  \centerline{(b)}
\end{minipage}
\caption{Evolution of opinions in the ER network with different rewiring probabilities at (a) $w=0$, and (b) $w=0.9$. Let $N=10^5$, $\langle k\rangle=20$, $d=0.1$ and $p=0.01$. The results show the average in 100 independent networks.}\label{fig:meaning}
\end{figure*}

Existing studies show that when a system does not reach a single global consensus but evolves to have multiple communities instead, the number of communities $M$ and tolerance $d$ roughly follow a relationship of $M\sim 1/2d$ [8, 23]. While such studies reveal the relationship between $M$ and $d$, to the best of our knowledge, none of them has investigated the relation between $M$ and mutation rate $p$. We show that, with other parameters given and fixed, different values of $p$ may lead to different numbers of communities at steady state.

To start, we first illustrate the relationship between $w$ and $M$. In Fig.~3, we show the results in an ER random network with $N=10^5$, $\langle k\rangle=20$, $d=0.1$ and $p=0.01$. We can see that when there is no rewiring operation, i.e., $w=0$, 4 communities will be formed up in the steady-state network; when $w=0.9$, however, 5 communities will be formed up in the steady-state network. In fact, it is a common conclusion applying to all the cases we have tested that a higher rewiring probability tends to lead to having more communities in the steady-state network.

Such an observation can be understood. To illustrate the dynamic process leading to the equilibrium, we show in Fig.~3 the opinion distribution in the two networks when the numbers of consensus formation operations (denoted as NCF in the figure) are $10^4$, $8\times10^4$ and $1.5\times10^5$ respectively, and at the final steady state, respectively. As discussed earlier, in both networks, there is an instability propagating inbound starting from the boundary. Specifically, two peaks form up close to the boundaries and gradually shift inbound. The inbound shifting however may be interrupted by rewiring operations since such operations, if fast enough, may make the two communities holding the outmost opinions quickly get largely separated from the rest part of the network. The opinion distribution of two communities shall then get largely stabilized, leaving a relatively larger opinion range in between, and consequently may allow more communities to be formed up in the further system evolution.

An example can be observed in Fig.~3. In Fig.~3(a), peaks 1 and 5 shift inbound until they reach $0.130$ and $0.870$ respectively, when NCF roughly equals $1.5\times10^5$. In Fig.~3(b), the inbound propagation is interrupted: the two communities holding the outmost opinions are largely stabilized when NCF roughly equals $8\times10^4$, with respective opinion peaks at $0.110$ and $0.890$. Such interruptions change the further co-evolution of opinion and network topology: the peaks of communities 2 and 3, after their emergence, propagate inbound and stop approximately at opinions $o=0.375$ and $o=0.625$ respectively in Fig.~3(a), which does not leave enough space for any further community to grow mature. For the network in Fig.~3(b), however, peaks 2 and 4 shift inbound and stop approximately at $o=0.305$ and $o=0.695$, respectively. Enough space is hence left for the central peak 3 to grow to full size. Note that such phenomena can be more easily observed for small values of $d$. For example, when $d=0.05$, by increasing $w$, we may observe two transitions in community number: first from 8 communities to 9 communities, and then from 9 to 10.

With the understanding that a large value of $w$ may lead to a larger number of opinion communities in the final-state networks, it would be interesting to study on the threshold value of $w$, denoted as $w_c$ hereafter, leading to the transition in the number of communities in different networks. Fig.~4(a) illustrates the threshold values in ER random networks with different average nodal degrees $\langle k\rangle$. For each $\langle k\rangle$ value, 10 random networks are generated and in each network, the threshold value $w_c$ is the average of 10 rounds of independent simulations. In each round of simulation, a trial-and-error approach is adopted to decide the value of $w_c$ with a step length of $0.01$. The figure shows the average in all the 10 networks with 95\% confidence range. It can be observed that in all these networks, a higher value of $w$ averagely leads to a steady state with more communities in the network; and the threshold value of phase transition in community number increases with the average nodal degree. This can be understood since a higher average nodal degree slows down the separation between different communities during the inbound instability propagation. Consequently it requests a higher rewiring rate to cause an increase in the number of communities at steady state

\begin{figure*}
\begin{minipage}{0.45\textwidth}
  \centerline{\includegraphics[width=8.0cm]{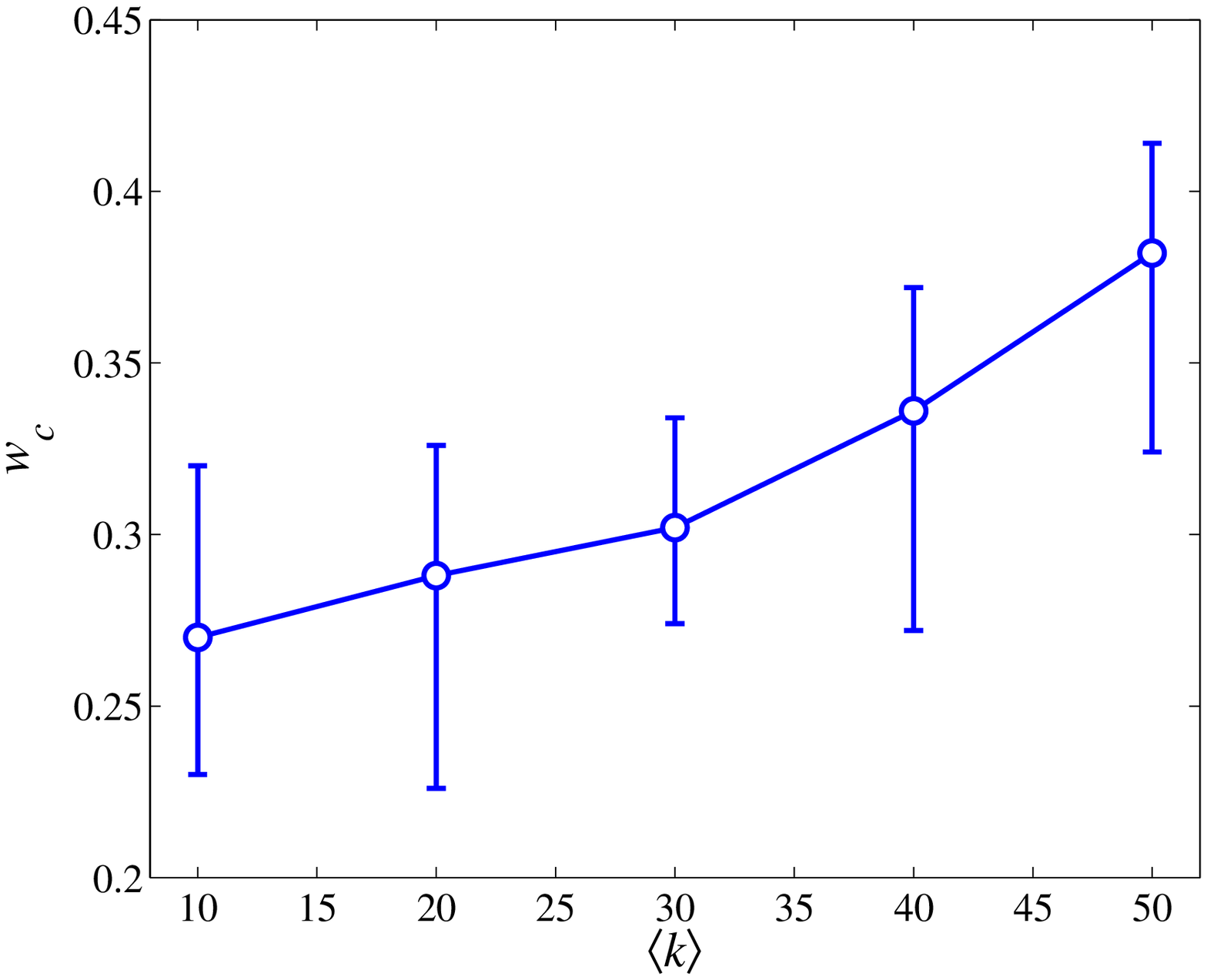}}
  \centerline{(a)}
\end{minipage}
\begin{minipage}{0.45\textwidth}
  \centerline{\includegraphics[width=8.0cm]{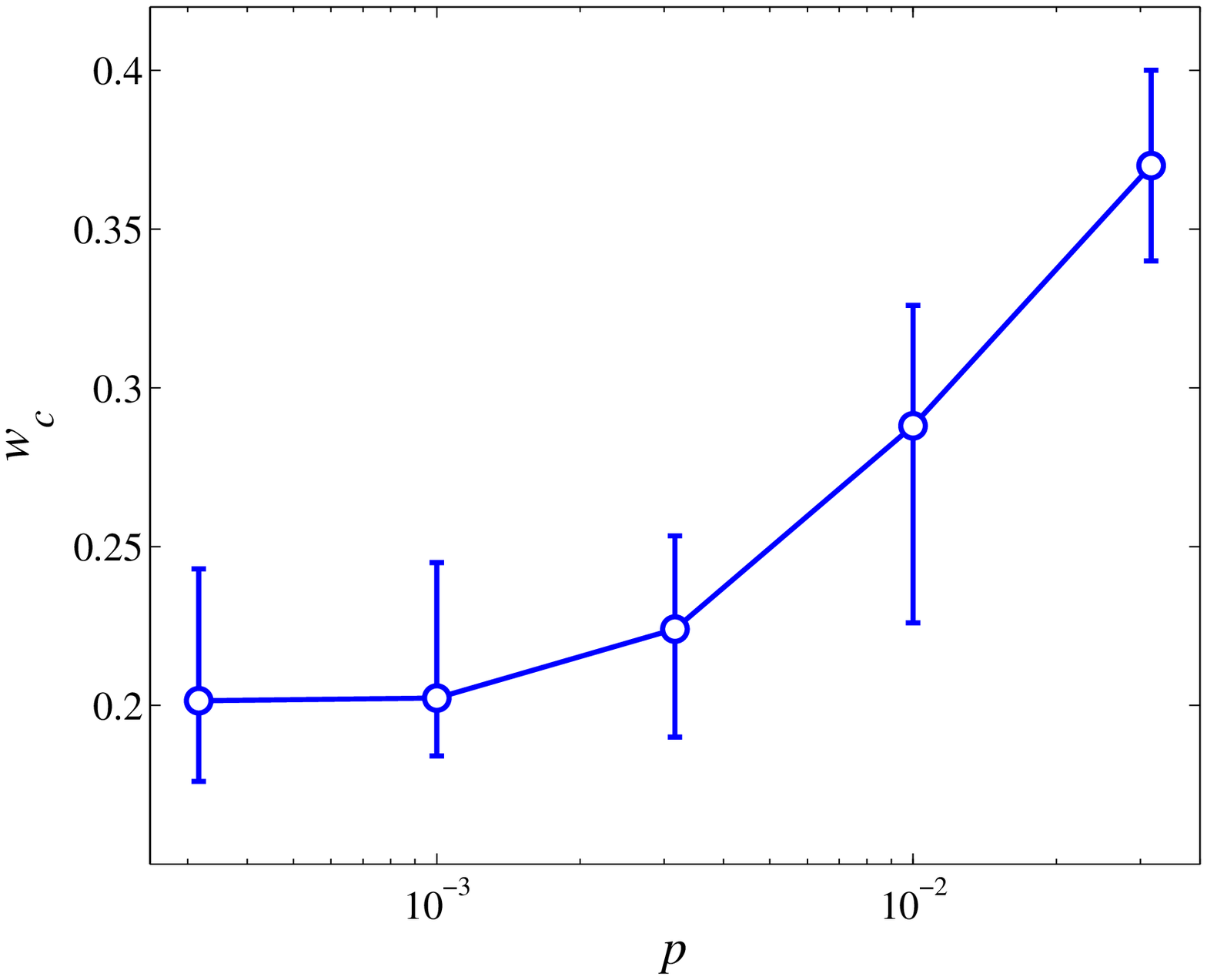}}
  \centerline{(b)}
\end{minipage}
\caption{Threshold value $w_c$ vs. (a) average degree $\langle k\rangle$, for $d=0.1$ and $p=0.01$; (b) mutation probability $p$, for $d=0.1$ and $\langle k\rangle=20$ in the ER random network. When $w>w_c$, 5 communities shall coexist in the steady-state network; while for $w<w_c$, only 4 communities would coexist. The 95\% confidence intervals are given for the average in 10 independent networks.}\label{fig:meaning}
\end{figure*}

Fig.~4(b) shows the relationship between the mutation probability $p$ and threshold value $w_c$. Again, 10 random networks are generated for each $p$ value, and in each network, the threshold value $w_c$ is the average of 10 rounds of independent simulations. The results show that a small value of $p$ (e.g., when $p$ increases from 0 to $1.0\times10^{-3}$) may not make any significant difference to the value of $w_c$. When the value of $p$ is large enough, however, a larger $p$ basically leads to a larger $w_c$. Such an observation is understandable: for a too small value of $p$, mutation's influences on the network evolution are trivial compared to those of consensus formation and rewiring, and thus make hardly any visible changes to $w_c$; while for a large enough value of $p$, random mutations essentially compromise the effects of rewiring in separating different communities, and hence increase the threshold value $w_c$. Equivalently we may draw the conclusion that for a given value of $w$, a larger value of $p$ tends to decrease the number of communities $M$ in the steady state. It is interesting, and somewhat counterintuitive, that a higher mutation rate, which we may expect to observe in a more tolerant or less stable society, actually decreases the number of opinion communities at steady state.

Note that the above observations are valid only when $p$ is large but not too large. A further increased $p$ value may further change the opinion distribution in the steady-state network, as we shall discuss in detail in Section~4.1.

\section{Theoretical analysis }\label{subsec:analysis}
In this section, we present theoretical analysis on the co-evolution of opinion and network structures. Specifically, a theoretical framework for describing the co-evolution process is proposed in Section~4.1, which matches well with the simulation results. In Section~4.2, we study in further details the impacts of a few factors on the modularity of the steady-state network.

\subsection{Analysis of the co-evolution process}\label{subsubsec:coevolution}
The co-evolution process and the final steady state can be mathematically described by mean-field analysis. The basic idea is to reveal the correlation between opinions on two end nodes of a randomly selected link. Due to consensus formation and rewiring operations, the two end nodes of each link tend to have similar opinions. Assume that time is continuous and nodes are selected at a rate $\epsilon$ to make decisions. Denote the probability density that a randomly selected node has an opinion $x_1$ and its random neighbor has an opinion $x_2$ at time $t$ as $P(t,x_1,x_2)$. We have $P(t,x_1,x_2 )=P(t,x_2,x_1)$. Further define a conditional probability density function $c(t,x_1,x_2)$, which denotes that at time $t$, given that a node has an opinion $x_1$, the probability that its random neighbor holds an opinion $x_2$. We have $c(t,x_1,x_2 )=P(t,x_1,x_2)/\rho(t,x_1)$, where $\rho(t,x_1)$ is the probability density of opinion $x_1$ at time $t$, i.e., $\rho(t,x_1)=\int_0^1{P(t,x_1,x)dx}=\int_0^1{P(t,x,x_1)dx}$. Assume that at each time step, a randomly selected node may mutate to another opinion $x$ with a probability density $b(x)$. For the simple random uniform mutation being considered in this article, $b(x)=1$ for any $x\in[0,1]$. The whole process can be described by the simple mean field analysis as follows.

Considering how mutation may help increase $P(t,x_1,x_2)$, we have:

\begin{equation}
\begin{split}
&increase_{mutation}(t,x_1,x_2)\\
&=\langle k\rangle \epsilon p\int_0^1{\rho(t,x)c(t,x,x_2)b(x_1)dx}\\
&~~~~+\langle k\rangle \epsilon p\int_0^1{\rho(t,x)c(t,x,x_1)b(x_2)dx}\\
&=\langle k\rangle \epsilon p\int_0^1{P(t,x,x_2)b(x_1)dx}\\
&~~~~+\langle k\rangle \epsilon p\int_0^1{P(t,x,x_1)b(x_2)dx}\\
&=\langle k\rangle \epsilon p\cdot \rho(t,x_2)+\epsilon \langle k\rangle p\cdot \rho(t,x_1).\\
\end{split}
\end{equation}
Meanwhile, mutation may decrease $P(t,x_1,x_2)$ at a rate of:
\begin{equation}
\begin{split}
&decrease_{mutation}(t,x_1,x_2)\\
&=\langle k\rangle \epsilon p\cdot \rho(t,x_1)c(t,x_1,x_2)\\
&~~~~+\langle k\rangle \epsilon p\cdot \rho(t,x_2)c(t,x_2,x_1)\\
&=\langle k\rangle \epsilon p\cdot P(t,x_1,x_2)\\
&~~~~+\langle k\rangle \epsilon p\cdot P(t,x_2,x_1).\\
\end{split}
\end{equation}
For each pair of nodes with an opinion difference less than $d$, they make consensus at a probability of $1-w$. Hence for $x_1=x_2$, consensus formation increases $P(t,x_1,x_2)$ at a rate of:
\begin{equation}
\begin{split}
&increase_{consensus1}(t,x_1,x_2)\\
&=\epsilon(1-w)\\
&~~~~\cdot \int_{-d/2}^{d/2}{\rho(t,x_1-x)c(t,x_1-x,x_2+x)dx}\\
&=\epsilon(1-w)\int_{-d/2}^{d/2}{P(t,x_1-x,x_2+x)dx},\\
\end{split}
\end{equation}
For $x_1\ne x_2$:
\begin{equation}
\begin{split}
&increase_{consensus2}(t,x_1,x_2)\\
&=(\langle k\rangle-1)\epsilon(1-w)\\
&~~~~\cdot \int_{-d/2}^{d/2}{P(t,x_1+x,x_1-x)c(t,x_1+x,x_2)dx}\\
&~~~~+(\langle k\rangle-1)\epsilon(1-w)\\
&~~~~\cdot \int_{-d/2}^{d/2}{P(t,x_2+x,x_2-x)c(t,x_2+x,x_1)dx}.\\
\end{split}
\end{equation}

Combining the above two different cases, we have:
\begin{equation}
\begin{split}
&increase_{consensus}(t,x_1,x_2)\\
&=\epsilon(1-w)\int_{-d/2}^{d/2}{P(t,x_1-x,x_2+x)dx}\cdot\delta_1(x_1,x_2)\\
&~~~~+(\langle k\rangle-1)\epsilon(1-w)\\
&~~~~\cdot \int_{-d/2}^{d/2}{P(t,x_1+x,x_1-x)c(t,x_1+x,x_2)dx}\\
&~~~~+(\langle k\rangle-1)\epsilon(1-w)\\
&~~~~\cdot \int_{-d/2}^{d/2}{P(t,x_2+x,x_2-x)c(t,x_2+x,x_1)dx},\\
\end{split}
\end{equation}
where $\delta_1(x_1,x_2)=1$ if $x_1=x_2$; otherwise $\delta(x_1,x_2)=0$.

Consensus formation decreases $P(t,x_1,x_2)$ at:
\begin{equation}
\begin{split}
&decrease_{consensus}(t,x_1,x_2)\\
&=\langle k\rangle\epsilon(1-w)\int_{-d}^d{P(t,x_1,x_1-x)c(t,x_1,x_2)dx}\\
&~~~~+\langle k\rangle\epsilon(1-w)\int_{-d}^d{P(t,x_2,x_2-x)c(t,x_2,x_1)dx}.\\
\end{split}
\end{equation}
For each pair of nodes with an opinion difference greater that $d$, rewiring happens at a probability of $w$. The link between the two nodes will be removed; and one of the end nodes of the link will be connected to a random node holding a similar opinion (with a difference less that $d$). Thus link rewiring increases $P(t,x_1,x_2)$ at a rate of:
\begin{equation}
\begin{split}
&increase_{rewiring}(t,x_1,x_2)\\
&=\epsilon w\int_d^1{P(t,x_1,x_1\pm x)\rho(t,x_2)/\phi(t,x_1,d)dx}\\
&~~~~\cdot\delta_2(x_1,x_2)\\
&~~~~+\epsilon w\int_d^1{P(t,x_2,x_2\pm x)\rho(t,x_1)/\phi(t,x_2,d)dx}\\
&~~~~\cdot\delta_2(x_1,x_2),
\end{split}
\end{equation}
where $\delta_2 (x_1,x_2)=1$ for $|x_1-x_2|\le d$; and $\delta_2(x_1,x_2)=0$ otherwise. Here we also define $\phi(t,x_1,d) =\int_{-d}^d{o(t,x_1+x)dx}$. Note that $\rho(t,x_1)/\phi(t,x_1,d)$ denotes the probability that at time $t$, a node with an opinion $x_2$ is chosen as the rewiring target and gets connected with a node with opinion $x_1$.

Similarly, it can be shown that link rewiring decreases $P(t,x_1,x_2)$ at a rate of:
\begin{equation}
\begin{split}
&decrease_{rewiring}(t,x_1,x_2)\\
&=\epsilon wP(t,x_1,x_2)(1-\delta_2(x_1,x_2))\\
&~~~~+\epsilon wP(t,x_2,x_1)(1-\delta_2(x_1,x_2)).
\end{split}
\end{equation}
Combining all the terms above, we have that the probability density function $P(t,x_1,x_2)$ changes at a rate of:
\begin{equation}
\begin{split}
&~\frac{dP(t,x_1,x_2)}{dt}\\
&=increase_{mutation}(t,x_1,x_2)\\
&~~~~+increase_{consensus}(t,x_1,x_2)\\
&~~~~+increase_{rewiring}(t,x_1,x_2)\\
&~~~~-decrease_{mutation} (t,x_1,x_2)\\
&~~~~-decrease_{consensus}(t,x_1,x_2)\\
&~~~~-decrease_{rewiring}(t,x_1,x_2)\\
\end{split}
\label{chap5:equ:5.10}
\end{equation}

\begin{figure*}[!hbt]
\begin{center}
\includegraphics[width=160mm]{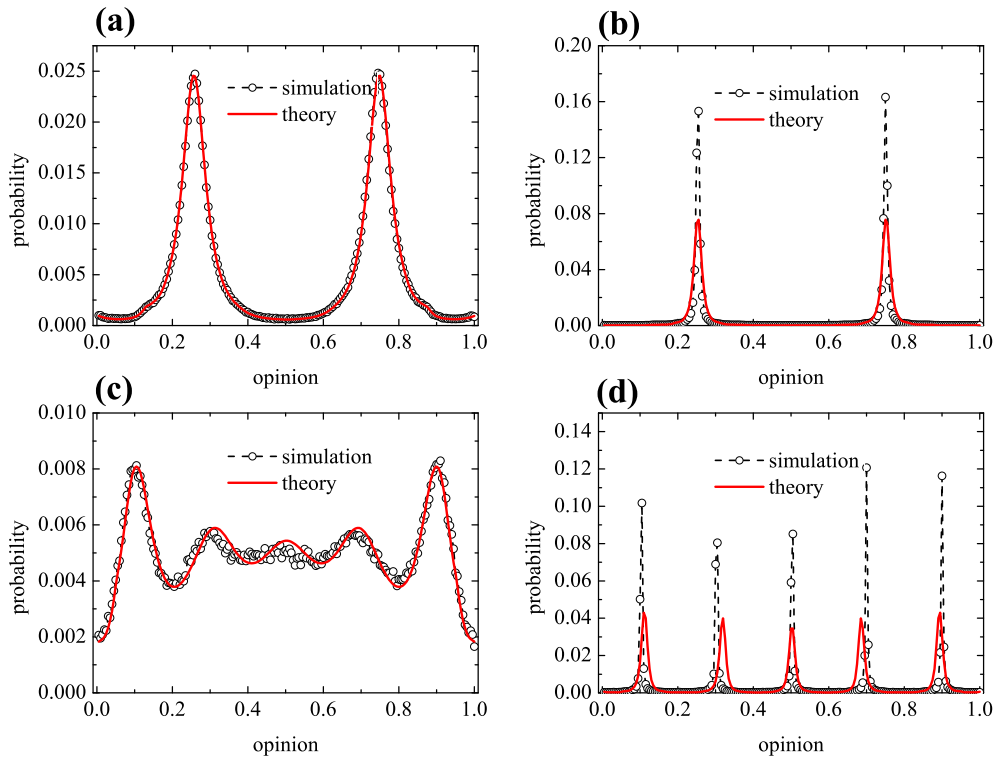}
\caption [An example of applying the MMI algorithm to allocate 2 monitors in an 8-node network.]{
\label{fig:fig3}{Comparison between simulation and theoretical results of steady-state opinion distribution in the ER random network with different sets of parameters: (a) $d=0.25$, $p=0.1$, $w=0.5$; (b) $d=0.25$, $p=0.01$, $w=0.5$; (c) $d=0.1$, $p=0.1$, $w=0.5$; and (d) $d=0.1$, $p=0.01$, $w=0.5$. We assume that the network size is $N=10^5$ and average nodal degree is $\langle k\rangle=20$.}}
\end{center}
\end{figure*}

\begin{figure*}[!hbt]
\begin{center}
\includegraphics[width=160mm]{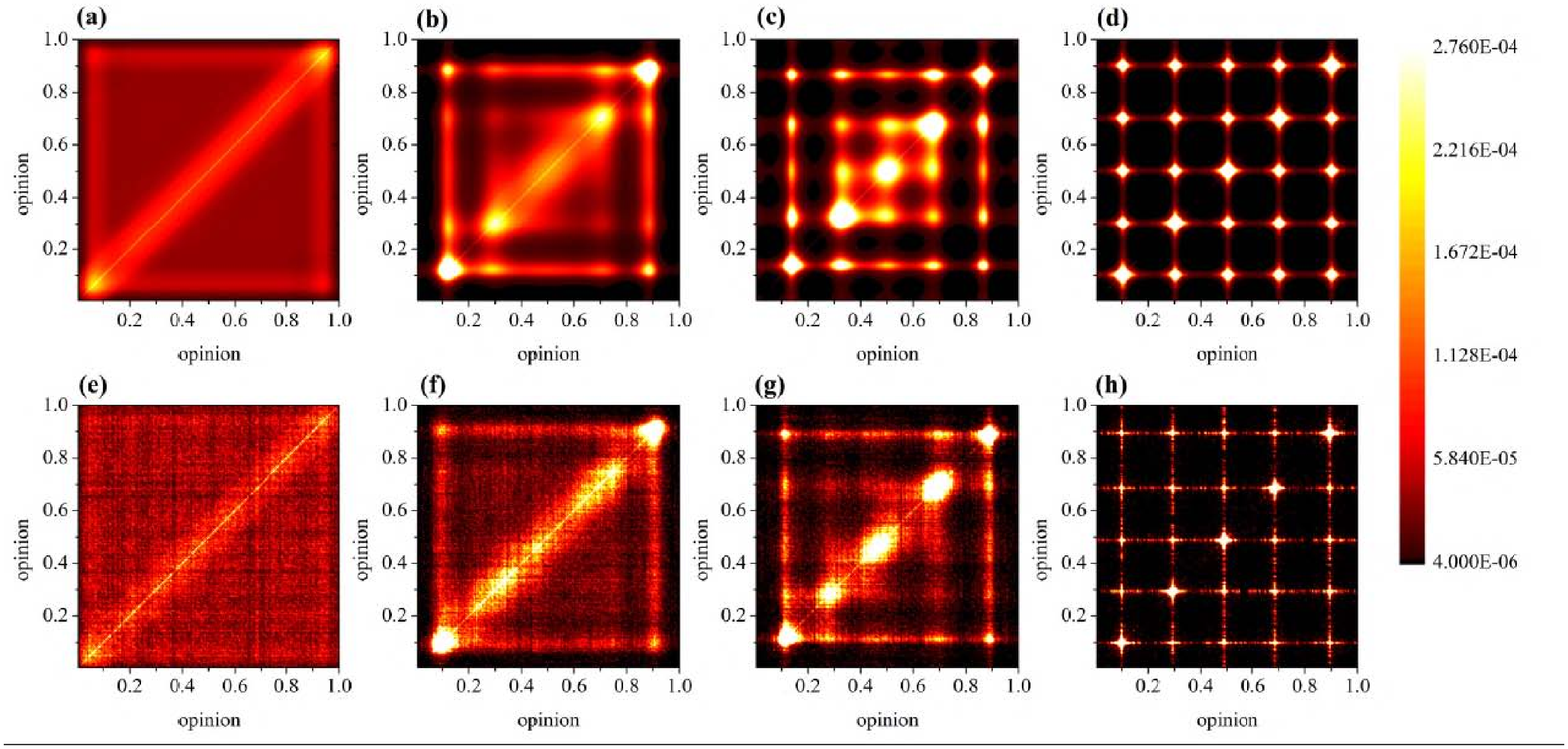}
\caption [An example of applying the MMI algorithm to allocate 2 monitors in an 8-node network.]{
\label{fig:fig4}{ Evolution of $P(t,x_1,x_2)$ for theoretical results at (a) $T=25$, (b) $T=100$, (c) $T=150$, (d) $T=5000$; and a single simulation case at time step (e) $T=25$, (f) $T=100$, (g) $T=150$, (h) $T=5000$. Note that for the number of time steps denoted as $t$, we have $t=2000T$.}}
\end{center}
\end{figure*}

The whole co-evolution process can be numerically calculated using Eqs.~(2)-(10). Extensive simulation results show that, for systems with a mutation probability $p>0$, with or without link rewiring, the initial opinion distribution basically has no influence on the final opinion distribution. Similar observations have been made on Deffuant model with noises (what we call mutation)~\cite{30}.

We first compare the theoretical results of the final-state opinion distribution with the simulation results. For each set of parameters, 100 ER random networks are generated, each with a size of $N=10^5$ and an average nodal degree of $\langle k\rangle=20$. Assume that one pair of nodes is selected at each time step. We record and average the opinion distribution every $2,000$ time steps from $t=8\times10^6$ to $t=10^7$ as the final state opinion distribution. It has been confirmed that the time is long enough for the network to reach the steady state. The average results in the 100 networks are considered as the steady-state opinion distribution corresponding to the given set of parameters.

Fig.~5 compares the simulation and theoretical results of opinion distribution in the steady-state network. The theoretical results are numerically calculated using Eq.~(10). We find that the theoretical results match well with the simulation results. The opinion distribution has a few bell-curve-shaped peaks with approximately equal``height". Note that the only exception is the case in Fig.~5(c) when $d=0.1$ and $p=0.1$. The relatively small opinion tolerance combined with the high mutation rate lead to a high density of inter-community links, which account for about 65\% of all the links. The high mutation rate weakens the community separation and growth to the level that communities 2, 3 and 4 never get a chance to be fully formed up. Even in the steady state, these three communities have a lower height than that of communities 1 and 5.

To achieve further insights into the co-evolution process, we examine the evolution of $P(t,x_1,x_2)$ in detail. Let $d=0.1$, $w=0.5$, $p=0.01$ and $\langle k\rangle=20$ where network would finally form into 5 communities. Fig.~6 shows both the theoretical and simulation results for the evolution of $P(t,x_1,x_2)$. The theoretical results are obtained from iterative numerical calculations of Eq.~(10) while setting $\epsilon=0.02$, which means that in each iteration, 2\% of the nodes make decisions. For the corresponding numerical simulations, we let each iteration contain $2,000$ time steps, which also allows about 2\% of the nodes make decision in each iteration. Denote the iteration number as $T$. We only show the results of a single round of simulation as the evolution speeds for different rounds of simulations vary due to noises, making it difficult to generate a clear image of average results. However, massive rounds of simulations have revealed that the system co-evolution always goes through nearly the same process, though not necessarily at the same speed; and there is a good match between theoretical and simulation results at different time as far as the pattern of $P(t,x_1,x_2)$ is concerned. Specifically, at the early stage of the co-evolution, nodes holding similar opinions gradually connect to each other at relatively higher probabilities. This is indicated by the high probability stripe along the diagonal of opinion distribution map as can be seen in Figs.~6(a) and 6(e). Then the boundary opinions (opinions close to 0 or 1) would make consensus first and gather together to form up the first two communities indicated by the two bright dots at the corner of Figs.~6(a) and 6(e). The formation of communities then propagates inbound, forming up two more communities, indicated by the two bright dots closer to the centers in Figs.~6(b) and 6(f). Finally, an opinion peak emerges around the central opinion value, as indicated by the bright dots in the center of Figs.~6(c) and 6(g). When the system reaches the final dynamic equilibrium, five communities are formed up, each of which mostly containing similar opinions around a peak value. Similar opinion holders connect to each other with a relatively high probability in these communities while they still preserve some connections with dissenters.

\begin{figure*}
\begin{minipage}{0.45\textwidth}
  \centerline{\includegraphics[width=8.0cm]{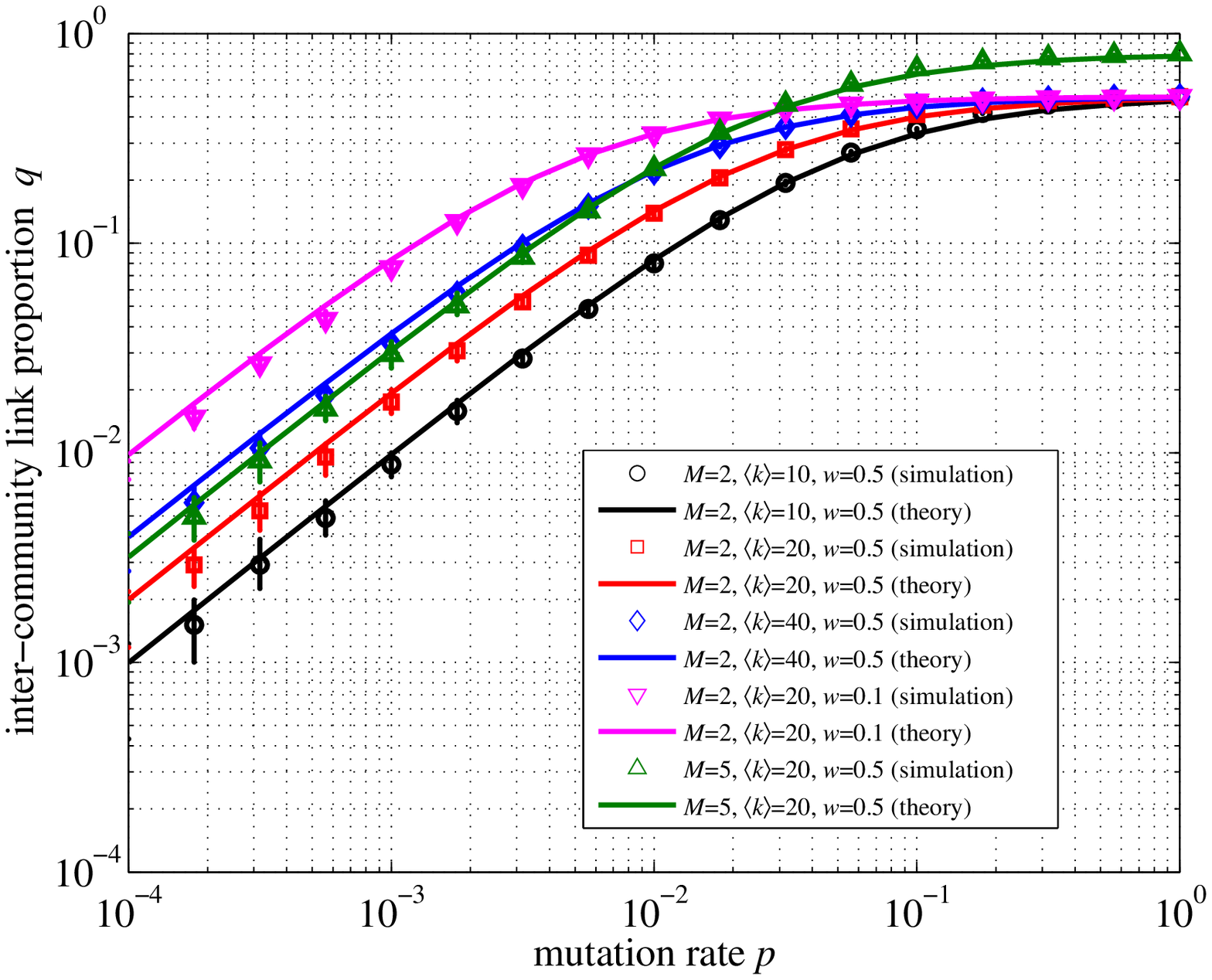}}
  \centerline{(a)}
\end{minipage}
\begin{minipage}{0.45\textwidth}
  \centerline{\includegraphics[width=8.0cm]{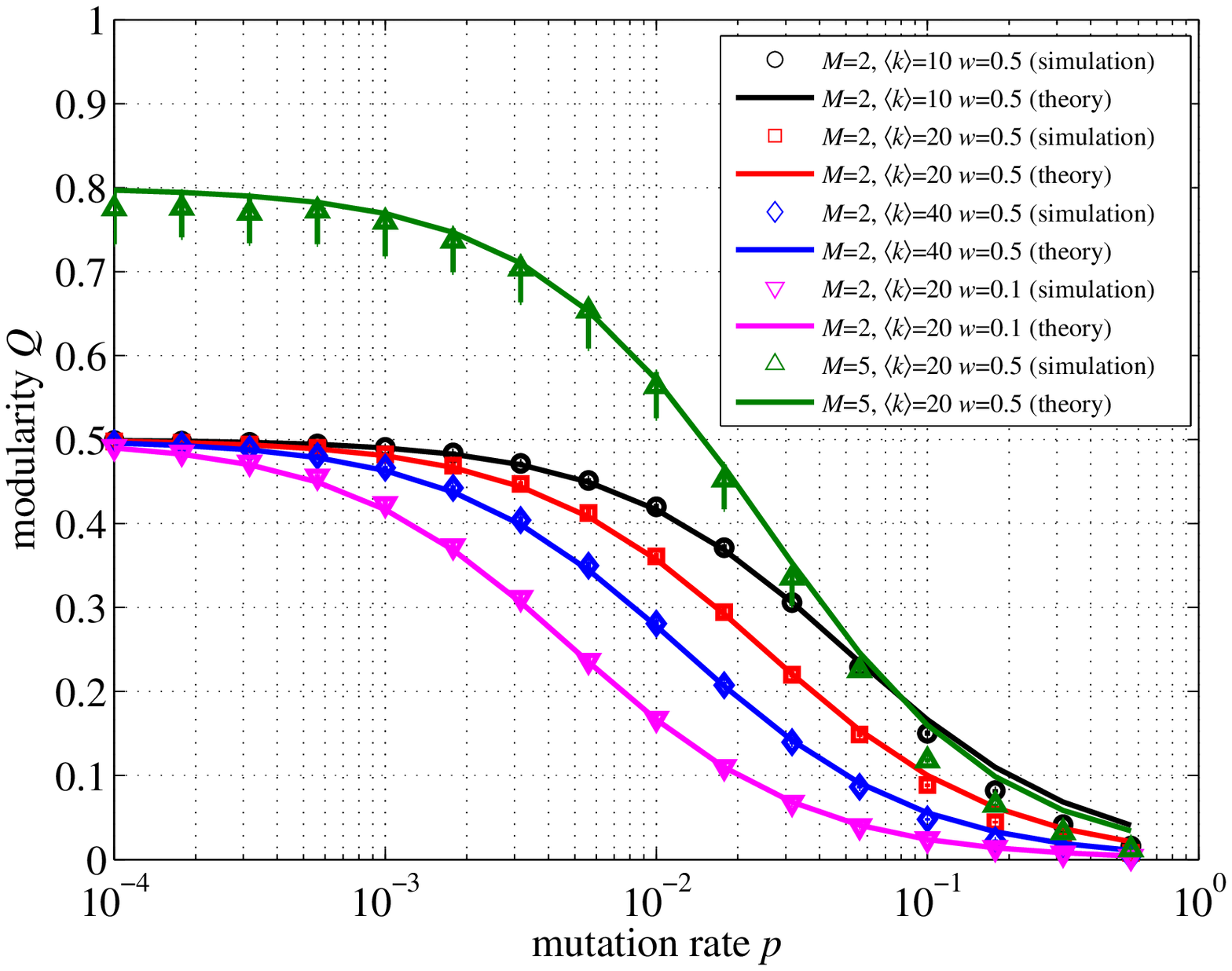}}
  \centerline{(b)}
\end{minipage}
\caption{(a) Proportion of inter-community links vs. mutation rate, and (b) Modularity vs. mutation rate. The 95\% confidence interval is given for the average results in 100 random networks.}\label{fig:meaning}
\end{figure*}

\subsection{Modularity of the steady-state networks}\label{subsec:community}
In this subsection, we further examine the impacts of a few factors, namely the rewiring speed, the mutation rate and the average nodal degree, on the modularity of the steady-state networks. From the discussions in Section~3, intuitively we may expect that since a higher mutation rate and a large average nodal degree tend to weaken the community structures, they may lead to a lower modularity level of the steady-state network. As to the rewiring speed, a higher rewiring speed tends to enhance community structures and consequently increase the network modularity at steady state. While such intuitive results can be confirmed by simulation as later we would see, it is of interest to study in detail how the network modularity is changed by the changes in these parameters, which is the main goal of this subsection.

We start by proposing an approximate analysis on the proportion of intercommunity links at steady state. To simplify the discussions, we assume that in the final state there are $M$ equal-size communities; that is, the $i^{th}$ community mainly contains those nodes holding opinions within the interval $I_i$  (defined in Section~3 that $I_1$ is $[0,1/M]$ and $I_i$ is $((i-1)/M,i/M]$ for $i=2,...,M$ respectively). This assumption is supported by the extensive simulation results reported in the last section, with the only exception for the case when the mutation rate is quite high. We argue that in most real-life systems, the mutation rate is of a low value, especially in those systems that can sustain a steady state. With such an assumption, the simple analysis can be derived as below.

Suppose that at the steady state, a randomly selected node averagely has $q$ proportion of its links being inter-community links. Let node pairs still be randomly selected at a rate of $\epsilon$. Due to mutation, the number of inter-community links is changed at a rate of:
\begin{equation}
\begin{split}
rate_{create}=\epsilon \left(p\langle k\rangle(1-q)-\frac{p\langle k\rangle q}{1-M}\right)\frac{M-1}{M}
\end{split}
\end{equation}

To explain Eq.~(11), we see that when a node with $q$ proportion of inter-community links and $(1-q)$ proportion of intra-community links is randomly selected and mutate to an opinion belonging to the range of a different community, all its intra-community links become inter-community links while averagely $1/(M-1)$ of its inter-community links become intra-community links. The absolute number of inter-community links therefore is changed by $\langle k\rangle p(1-q)- \langle k\rangle pq/(1-M)$. The probability that a mutation does change a randomly selected node's opinion to be within the opinion range of another community is $(M-1)/M$.

At the same time, due to the rewiring operation, inter-community links are broken at a rate of:
\begin{equation}rate_{rewiring}=\epsilon qw\end{equation}		
This equation shows that for a randomly chosen node and its random neighbor, the link connecting them has a probability $q$ to be an inter-community link; and if it is an inter-community link, it is broken at a probability $w$. Note that here we neglect the effects of consensus formation as we assume that in the final steady state, consensus formation almost only happens between nodes in the same community. The dynamics of the inter-community links at steady state therefore can be expressed as:
\begin{equation}
\begin{split}
\epsilon qw=\epsilon \left(p\langle k\rangle(1-q)-\frac{p\langle k\rangle q}{1-M}\right)\frac{M-1}{M}
\end{split}
\end{equation}

And the inter-community link proportion q can be expressed as:
\begin{equation}
\begin{split}
q=\frac{p\langle k\rangle}{w+p\langle k\rangle}\frac{M-1}{M}
\end{split}
\end{equation}

With Eq.~(14), we can then proceed to calculate the modularity of the network. We adopt the definition in \cite{37} that the network modularity can be calculated as:
\begin{equation}
\begin{split}
Q=\frac{1}{4m}\sum_{ij}{\frac{A_{ij}-(k_ik_j)}{2m}s_is_j}
\end{split}
\end{equation}
where $A$ is the adjacency matrix, $m$ is the number of edges and $k_i$ is the nodal degree of node $i$. For all the possible combinations of $i$ and $j$, if nodes $i$ and $j$ are in different communities, $s_is_j+1=0$; and $s_is_j+1=2$ otherwise. Since a randomly selected node averagely has $q$ proportion of inter-community links, for a given node $i$ and a randomly selected node $j$, the probability that they are connected and in the same community is $k_i(1-q)/N$, while with a probability $1/M$, the two nodes are in the same community. Therefore there exists simple relationship between $Q$ and $q$ that:
\begin{equation}
\begin{split}
&Q=\frac{1}{4m}\sum_i{\left(k_i-\frac{k_i2m/M}{2m}\right)}(1-q)\\
&~~~~-\frac{1}{4m}\sum_i{\left(k_i-\frac{k_i2m(M-1)/M}{2m}\right)}q\\
&=\frac{1}{2m}\sum_i{\left(k_i \frac{M-1}{M}-k_iq\right)}~~~~~~~~\\
&=\frac{M-1}{M}-q~~~~~~~~~~~~~~~~~~~~~~~~~\\
\end{split}
\end{equation}

To verify the analysis, we carry out simulations on the ER random networks with $N=10^5$ and $\langle k\rangle=10,20,40$ respectively. We adopt parameter values $d=0.25$, $w=0.1$ and $0.5$ where the networks will finally have 2 communities, and $d=0.1$, $w=0.5$ where the networks will finally have 5 communities. Different opinions are uniformly distributed at the beginning. We examine different cases with different mutation rates of $p=10^{(-4+n/4)}$ for $n$ from 0 to 16. For each parameter set, we generate 100 random networks. And in each time step, we let a single pair of nodes be randomly selected. We record the proportion of inter-community links and calculate the network modularity every 2000 time steps from $t=8\times10^6$ to $t=10^7$; and finally average the results of these 1000 calculations.

Fig.~7(a) shows the change of inter-community link proportion $q$ with the mutation rate $p$. We also evaluate the effects of rewiring speed $w$ and the average degree $\langle k\rangle$ on $q$. We see that in all the different cases, theoretical analyses match well with simulation results. Notice that when the mutation rate $p$ is very low, we have that $\langle k\rangle p\ll w$; from Eq.~(14) we have $w+\langle k\rangle p\approx w$ and hence $q\sim p$. The proportion of inter-community links therefore increases approximately linearly with the mutation rate. This matches the observation in Fig.~7(a) that in a log-log scale plot, the relation between $p$ and $q$ approximates a linear function with a slop value 1 when $p$ is small enough. The increase of $q$ becomes slower when the mutation rate $p$ is high, which can be derived from Eq.~(14) and also can be easily understood: as aforementioned, a higher mutation rate leads to a larger proportion of inter-community links and a smaller proportion of intra-community links. When a mutation operation changes the opinion of a node to be within the range of another community, all the node's intra-community links become inter-community links while a part of their inter-community links become intra-community links. Such mutations become less effective in increasing the number of inter-community links when there is already a large proportion of inter-community links. The increasing speed of $q$ thus becomes slower than a linear function of $p$.

Rewiring rate $w$ also has its impacts on the inter-community link proportion $q$. A larger value of $w$ basically leads to a smaller value of $q$ when other parameters are fixed. The reason is simple: a higher rewiring speed leads to a smaller number of inter-community links. While such an observation is obvious, it is however interesting to observe from (14), and Fig.~7(a) as well, that when $p$ is small enough, for given $p$ and $\langle k\rangle$, $q$ increases approximately linearly proportional to $1/w$.

As to the effects of $\langle k\rangle$ on $q$, it can be easily derived from Eq.~(14) that when $\langle k\rangle p\ll w$, $q$ increases almost linearly with $\langle k\rangle$. Note that the steady state is a dynamic equilibrium where the speed of increasing inter-community links by opinion mutation statistically speaking equals the speed of removing inter-community links by rewiring. A high average nodal degree allows the number of inter-community links to be increased faster by a given number of mutation operations (i.e., a given $p$). It thus takes a higher proportion of inter-community links at steady state to allow a larger number of inter-community link removals such that the dynamic balance can be achieved. The effects again become less significant when $p$ is large, due to the same reason as we discussed earlier that mutation becomes less effective in increasing the number of intercommunity links under such case.

With the understanding of the relation between $q$ and a few factors, effects of these factors on the system modularity value $Q$ can be easily derived from Eq.~(16): when $p$ is very small, $Q$ decreases approximately linearly with $p$, $\langle k\rangle$ and $1/w$ . The decreasing speed becomes slower when $p$ gets larger. To allow better observation of modularity values within the range that we are interested, Fig.~7(b) plots the results in log-linear scale. Note that it is confirmed that $Q$ decreases approximately linearly with $p$ when $p$ is small though it does not appear to be an obvious observation in the figure. Further, it is interesting to observe that when $w$ is of a high value, e.g., at 0.5, it takes a relatively high value of $p$ (higher than $10^{-3}$) to let $Q$ significantly shift away from $(M-1)/M$. For a smaller value of $w$, e.g., at 0.1, a low value of $p$ can already lead to nontrivial changes to $Q$. This matches the real-life experiences that in a more tolerant social system (with a lower value of $w$), opinion changes are more effective in increasing the number of interconnections between different opinion communities.

\section{Conclusions}\label{sec:conclusion}
In this work, we evaluated the complex co-evolution process of opinions and system structures where three different factors, namely opinion formation, link rewiring and opinion mutation are integrated together into the same model. It was observed that such a model would allow system to evolve into a dynamic equilibrium with multiple communities, each of which holding a range of opinions with a bell-curve-style distribution and with non-trivial intercommunity links in between. Such a system, as we claim, better resembles the observations in the real life and would allow system to easily further evolve when internal/external conditions change. It is also observed that a few different factors, rather than tolerance between different opinions alone, may contribute to deciding the number of communities in the final-state networks. An analytical framework was proposed to describe the co-evolution process with satisfactory precision. Further, the relation between system co-evolution and system modularity was carefully studied. It was revealed that there exists a linear relationship between system modularity and a few factors when the mutation rate is low. Our study shall help better understand how different factors work together leading to the complex co-evolution dynamics that we may observe in many complex social systems.
It is revealed in this article that different mutation rates may lead to different final steady states. In our preliminary studies \cite{38, 39}, it has been further observed that the effects of different mutation patterns may be equally, if not more, significant. Further studies are needed to fully understand the effects of mutation in opinion formation on complex social networks.

\section*{Acknowledgement}
The work is partially supported by Ministry of Education, Singapore, under contracts RG28/14, MOE2014-T2-1-028 and MOE2016-T2-1-119, respectively.

\bibliographystyle{unsrt}

%\bibliography{monitor}

%%%%%%%%%%%%%%%%%%%%%%%%%%%%%%%%%%%%%%%%%%%%%%%%%%%%%%%%%%%%%%%%%%%%%%%%%%%%%
% Place all of the references you used to write this paper in a file
% with the same name as following the \bibliography command
%%%%%%%%%%%%%%%%%%%%%%%%%%%%%%%%%%%%%%%%%%%%%%%%%%%%%%%%%%%%%%%%%%%%%%%%%%%%%

%\bibliography{sample-paper}

%\bibliographystyle{prsty}
%\begin{thebibliography}{99}
%\bibitem{melissinos1966}Melissinos, A.C., Experiments in Modern
%  Physics - 1st Edition, Academic Press,  [1966]
%\bibitem{melissinos2003}Melissinos, A.C., Napolitano, J.,  Experiments in Modern
%  Physics - 2nd Edition, Academic Press,  [2003]
%\bibitem{bevington2003}Bevington and Robinson, Data Reduction and
%  Error Analysis for the Physical Sciences - 3rd Edition, McGraw-Hill,
%  [2003]
%\bibitem{pritchard1990}Professor D. Pritchard, Personal Communication
%\end{thebibliography}

%%%%%%%%%%%%%%%%%%%%%%%%%%%%%%%%%%%%%%%%%%%%%%%%%%%%%%%%%%%%%%%%%%%%%%%%%%%%%
%\begin{acknowledgments} FAC gratefully acknowledges Dr. Francine Brown for
%her early reviews of this manuscript.
%\end{acknowledgments}

%%%%%%%%%%%%%%%%%%%%%%%%%%%%%%%%%%%%%%%%%%%%%%%%%%%%%%%%%%%%%%%%%%%%%%%%%%%%%
\clearpage

\end{document}